\documentclass[doublecol]{epl2}

\newcommand{\ff}[1]{{\bm #1}}
\newcommand{\be}{\begin{equation}}
\newcommand{\ee}{\end{equation}}

\title{First order Mott transition at zero temperature in two dimensions: Variational plaquette study}

\shorttitle{First order Mott transition at zero temperature in two dimensions}

\author{
Matthias Balzer, \inst{1}
Bumsoo Kyung, \inst{2}
David S\'en\'echal, \inst{2}
A.-M. S. Tremblay, \inst{2}
Michael Potthoff \inst{3}
}

\institute{
  \inst{1} Institut f\"ur Theoretische Physik und Astrophysik, Universit\"at W\"urzburg, Germany \\
  \inst{2} D\'epartement de physique and RQMP, Universit\'e de Sherbrooke, Qu\'ebec, Canada J1K 2R1 \\
  \inst{3} I.\ Institut f\"ur Theoretische Physik, Universit\"at Hamburg, Germany
}


\abstract{ 
The nature of the metal-insulator Mott transition at zero temperature has been discussed for a number of years. Whether it occurs through a quantum critical point or through a first order transition is expected to profoundly influence the nature of the finite temperature phase diagram. In this paper, we study the zero temperature Mott transition in the two-dimensional Hubbard model on the square lattice with the variational cluster approximation. This takes into account the influence of antiferromagnetic short-range correlations. By contrast to single-site dynamical mean-field theory, the transition turns out to be first order even at zero temperature.
}

\begin{document}

\maketitle

\section{Introduction}

The correlation-driven transition from a paramagnetic normal Fermi liquid at
weak coupling to a paramagnetic Mott insulator at strong coupling is one of
the most important paradigms in solid-state theory \cite{Mot49,Mot90}. For
example, the Mott state is suggested to represent the proper starting point
for theoretical studies of the extremely rich and difficult correlation
physics of two-dimensional systems such as cuprate-based high-temperature
superconductors \cite{And87}. A big step forward in the understanding of the
Mott transition was made by applying the dynamical mean-field theory (DMFT)
\cite{MV89,GKKR96,KV04} to the single-band Hubbard model which is believed
to capture the main physics of the Mott transition in a prototypical way.
The Hamiltonian is given by
\begin{equation}
H = - t \sum_{\langle ij \rangle,\sigma} c_{i\sigma}^\dagger c_{j\sigma} + U
\sum_i n_{i\uparrow} n_{i\downarrow}  
\label{eq:hub}
\end{equation}
where the summation is over site indices $i,j$, and $\sigma$ is the spin label.
$c_{i,\sigma }^\dagger$ and $c_{i,\sigma }$ are the particle creation and 
annihilation operators and $n_{i,\sigma }=c_{i\sigma}^\dagger c_{i\sigma}$.
Each doubly occupied site costs an energy $U$.
All the numerical results are presented in units where $t=1$.

A number of objections concerning the DMFT picture of the Mott
transition have been raised~\cite{N98} and answered \cite{K99}. However, the
main criticism comes from the neglect of the feedback of non-local magnetic 
correlations on the single-particle dynamics. 
This leads to a description of the paramagnetic Mott insulator with a
macroscopically large ground-state entropy of $S = L \ln 2$ (where $L \to
\infty$ is the number of sites). On the level of one-particle excitations,
superexchange does not lift the $2^L$-fold degeneracy arising from the spins
of the localized electrons. This must be considered as a mean-field artifact
which has profound consequences for the phase diagram: At finite
temperatures, $T>0$, the large entropy term stabilizes the Mott insulator in
a situation where it competes with a metallic Fermi liquid.

Cluster extensions of the DMFT \cite{MJPH05} can cure this defect since they
incorporate the feedback of \emph{short-range} antiferromagnetic
correlations. This has motivated a number of previous studies which put the
relevance of the mean-field picture for the transition into question and
focus on the qualitative change of the phase diagram due to short-range
magnetic correlations in, say, two dimensions 
\cite{MDR99,KT06,ZI07,OMTK08,NSST08,PHK08,GWTM08}.

\begin{figure}[tbp]
\includegraphics[width=0.99\columnwidth]{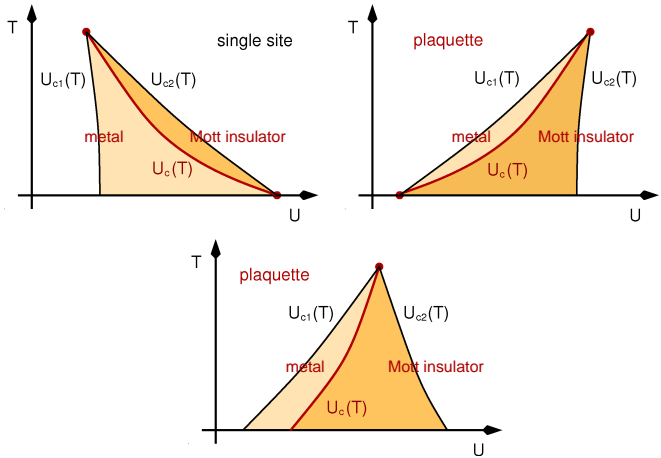}
\caption{ \emph{Top left:} sketch of the DMFT phase diagram. Coexistence of
a metallic and a Mott insulating solution is found in the yellow region. A
first-order Mott transition takes place at $U_c(T)$ (red line). The
first-order line ends in second-order critical points (red dots) at $T=T_c$
and $T=0$. \emph{Top right:} possible phase diagram within plaquette DMFT. The
first-order line ends in $U_{c1}$ for $T=0$. \emph{Bottom:} the phase diagram
supported by our plaquette VCA calculations. The first-order line does not
end in a critical point at $T=0$. }
\label{fig:fig1}
\end{figure}

Recently, novel quantum Monte-Carlo techniques to solve the problem for a
plaquette of correlated sites have been used to study the transition at
finite $T$ on the square lattice with nearest-neighbor hopping $t$. While
different embeddings of the plaquette were considered, namely the cellular
DMFT (C-DMFT) \cite{PHK08} and the dynamical cluster approximation (DCA)
\cite{GWTM08}, it turned out that salient features of the DMFT phase diagram
\cite{GKKR96} are preserved (see fig.\ref{fig:fig1}). In particular, there
is again coexistence of a metallic and an insulating phase in a certain $U-T$
range at half-filling bounded by lines $U_{c1}(T)$ and $U_{c2}(T)$.
Comparison of their respective free energies leads to a first-order
transition line $U_c(T)$ which, at a temperature $T_c$, ends in a
second-order critical point above which there is a smooth crossover only.

For $T<T_c$ there are different possibilities: Within single-site
DMFT (fig.~\ref{fig:fig1}, top left), the insulator wins at higher
temperatures due to the residual high entropy of the insulator. This implies
that $U_c(T)$ is increasing with decreasing $T$. Consequently, the line ends
for $T=0$ in another second-order critical point which must coincide with
the point up to which a metallic solution can be found, i.e.\ $U_c(0)=U_{c2}(0)$. 
At $T=0$ the metal is stable in the entire coexistence
region. Contrary, within a plaquette DMFT, the entropy is low in the
insulator due to short-range singlet formation. The phase diagram fig.~\ref{fig:fig1} 
(top right) is obtained if the metal always wins at higher $T$.
As compared with single-site DMFT, the critical line $U_c(T)$ bends back,
and $U_c$ \emph{decreases} with decreasing $T$ \cite{PHK08}. Hence, 
$U_c(0)=U_{c1}(0)$, and at $T=0$ the insulator is stable in the entire
coexistence region.

\section{Main results}

Here we use the variational cluster approximation (VCA) \cite{Pot03a,PAD03}
to embed
a plaquette of four correlated sites and four or eight
uncorrelated bath sites in the lattice. As the method is thermodynamically consistent and
focuses on the optimization of a thermodynamical potential, it is ideally
suited to distinguish between different phase diagram topologies. Here we
consider $T=0$ using the Lanczos method as a cluster solver. Our results are
consistent with the previous plaquette DMFT studies \cite{PHK08,GWTM08} but
support yet another low-temperature phase diagram (fig.~\ref{fig:fig1}, bottom). 
We find a sizable interaction range $U_{c1} < U < U_{c2}
$ where the metallic and the insulating solution are coexisting at zero
temperature. The $T=0$ endpoint of the line of first-order transitions 
$U_{c}=U_c(T=0)$
does neither coincide with $U_{c1}$ nor with $U_{c2}$, so that the Mott
transition is discontinuous also at $T=0$.

\begin{figure}[tbp]
\centering \includegraphics[width=0.8\columnwidth]{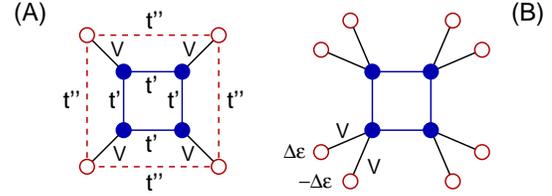}
\caption{Reference systems consisting of four correlated sites
with $U\ne0$ (filled, blue) and four (A) or eight (B)
uncorrelated sites with $U=0$ (open, red). One-particle variational
parameters: ``hybridization'' $V$, hopping between correlated ($t^{\prime}$)
and uncorrelated sites ($t^{\prime\prime}$) and shift ($\pm \Delta
\varepsilon$) of the energies of the bath sites with respect to $\mu$. Here,
$\mu$ is the common chemical potential for both cluster and bath. For
arbitrary one-particle parameters a space of trial self-energies $\ff
\Sigma$ is spanned on which a stationary point of the SFT grand potential 
$\Omega[{\bm \Sigma}]$ is searched.}
\label{fig:fig2}
\end{figure}

\section{Method}

Using a plaquette of four sites to generate an approximate self-energy 
$\Sigma_{ij}(\omega)$ for the infinite $D=2$ square lattice, represents the
essential step to go beyond the single-site DMFT approximation 
$\Sigma_{ij}(\omega) \approx \delta_{ij}\Sigma(\omega)$. An in principle
ideal embedding of the cluster in the infinite lattice could only be
achieved with a continuum of bath degrees of freedom (uncorrelated sites
with $U=0$) attached to the correlated four-site cluster in the spirit of
quantum-cluster theories \cite{MJPH05}. For $T=0$, however, this is not yet
accessible with presently known cluster solvers. As far as static quantities
and the thermodynamical phase diagram are concerned, however, it is
fortunately well known that a few bath sites can be sufficient for reliable
predictions \cite{CK94,Pot03b,BHP08,KSG08}. This holds on the single-site
level where the DMFT phase diagram for the Mott transition can be recovered
qualitatively with a single bath site only \cite{Pot03b} as well as for
cluster approximations, as has been demonstrated recently in one dimension
for the filling-controlled transition \cite{BHP08,KSG08}.

\begin{figure}[tbp]
\centering \includegraphics[width=0.95\columnwidth]{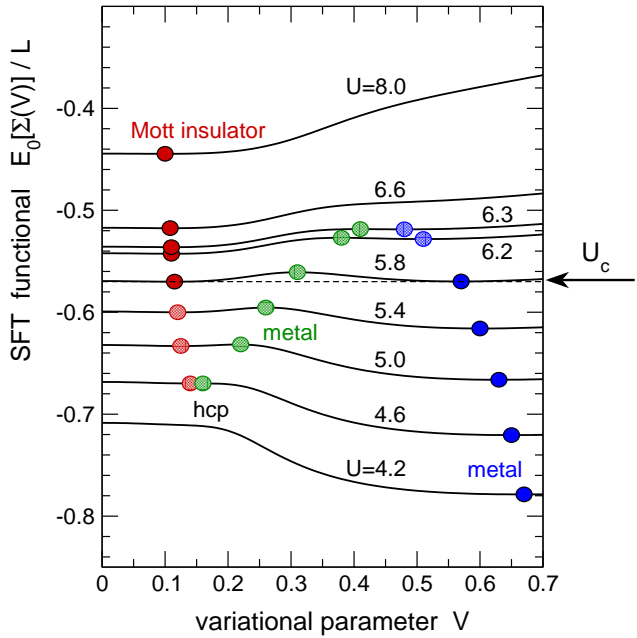}
\caption{ 
SFT grand potential shifted by the constant $\protect\mu N$, $E_0[{
\bm \Sigma}] = \Omega[{\bm \Sigma}] + \protect\mu N$, and evaluated for
trial self-energies ${\bm \Sigma}_{V,0,0}$ (see fig.~\protect\ref{fig:fig2}
(A)) as a function of $V$. Results for different $U$ at $T=0$ and
for half-filling $N=L$ (chemical potential $\protect\mu = U/2$). Circles
indicate stationary points. Here, the value of the functional equals the
ground-state energy $E_0$ (per site). Filled circles: stable metallic (blue)
and insulating (red) phase. Shaded circles: metastable phases. At $U_c$
(arrow) there is a discontinuous metal-insulator transition. A third
(metastable) metallic phase (green circles) continuously coalesces with the
insulating one at a hidden (metastable) critical point (hcp). The energy
unit is fixed by setting $t=1$.
}
\label{fig:fig3}
\end{figure}

To address the phase diagram of the Mott transition, we have to employ a
\emph{thermodynamically consistent} method to optimize the cluster
self-energy and to fix the plaquette and bath parameters. For the reference
systems displayed in fig.~\ref{fig:fig2} this can be achieved
within the self-energy-functional theory (SFT). The SFT has been described
in detail in refs.\ \cite{Pot03a,Pot03b,PAD03}. The main idea is to use the
reference system for spanning a space of trial self-energies.
Referring to the reference system in fig.~\ref{fig:fig2} (A), for example, the trial self-energy is parametrized as $\ff \Sigma = \ff \Sigma_{V,t',t''}$ and calculated at a given $U$ and at $T=0$ for each set $(V,t',t'')$ from the cluster Green's function $\ff G'$ via Dyson's equation.
On this space, the general SFT grand potential as a functional of the
self-energy, $\Omega[{\bm \Sigma}]$, can be evaluated exactly via
\begin{equation}
\Omega[{\bm \Sigma}] = \Omega^{\prime}+ \mbox{Tr} \ln {\bm G } - \mbox{Tr}
\ln {\bm G}^{\prime}\: .
\end{equation}
Here $\Omega^{\prime}$ is the grand potential of the reference system and 
${\bm G}$ is the lattice Green's function obtained via 
${\bm G } = 1 / ({{\bm G}_{0}^{-1} - {\bm \Sigma}})$ from the free lattice 
Green's function ${\bm G}_0$. A physical state is found as a stationary point 
of $\Omega[{\bm \Sigma}_{V,t^{\prime},t^{\prime\prime}}]$ as a function of the variational
parameters $(V,t^{\prime},t^{\prime\prime})$. To this end we exploit the
particle-hole symmetry of the Hubbard model (\ref{eq:hub}) at half-filling,
use the (band) Lanczos method for a simultaneous calculation of all elements
of the cluster Green's function ${\bm G}^{\prime}$, and employ the $Q$-matrix technique
or integration along the imaginary frequency axis to evaluate $\mbox{Tr} \ln \ff G / \ff G'$ \cite{trlng}.
In addition, we make use of spatial symmetries to limit the number of
independent one-particle parameters to a minimum.

\section{Parameter optimization}

For a simultaneous independent optimization of $V$, $t^{\prime}$ and 
$t^{\prime\prime}$ (reference system fig.~\ref{fig:fig2} (A)) we
apply the downhill simplex method and/or iterative one-dimensional
optimizations to find local minima of 
$|\nabla \Omega[{\bm \Sigma}_{V,t^{\prime},t^{\prime\prime}}] |^2$ from which (if there are more than
one) only those are retained for which 
$\Omega[{\bm \Sigma}_{V,t^{\prime},t^{\prime\prime}}]$ has a vanishing gradient. Particle-hole
symmetry fixes the optimal values $\varepsilon_{c,\mathrm{opt}} = 0$ and 
$\varepsilon_{b,\mathrm{opt}} = \mu = U/2$ for the on-site energies of the
correlated and the bath sites, respectively. It has also been checked 
\emph{numerically} that the SFT functional is stationary at these values. For the
optimal hopping between bath sites we find $|t^{\prime\prime}_{\mathrm{opt}}| < t/25$ 
in the entire $U$ range considered. Note that for a finite number
of bath sites $L_b$ the inclusion of $t^{\prime\prime}$ enlarges the space
of trial self-energies while for continuous baths ($L_b\to\infty$) the
approximation becomes equivalent \cite{PAD03} with C-DMFT where a coupling
of baths attached to different correlated sites is not needed for this lattice
geometry.

The optimal hopping parameter between the correlated sites turns out as 
$t^{\prime}_{\mathrm{opt}} = t + \Delta t^{\prime}_{\mathrm{opt}}$ with a
small positive $\Delta t^{\prime}_{\mathrm{opt}} <t/10$ for the $U$ range
considered here. A considerably larger $\Delta t^{\prime}_{\mathrm{opt}}$ is
only found in the limit $U \to 0$. For more itinerant electrons a stronger
enhancement of the \emph{intra}cluster hopping is needed to (partially)
compensate for switching off the \emph{inter}cluster hopping in the
approximation for the VCA self-energy. Note that $t^{\prime}_{\mathrm{opt}}=t
$ for $L_b \to \infty$. This is easily derived by a $1/\omega$ expansion of
the SFT Euler equation (equivalently the C-DMFT self-consistency equation).
The optimization of $t^{\prime}$ and $t^{\prime\prime}$ actually turns out
to be almost irrelevant as compared to $V$. Setting $t^{\prime}=t$ and 
$t^{\prime\prime}=0$ and performing a one-dimensional optimization of $V$
only, leads to changes in the optimal $V$ of less than 1\%. The change in
the ground-state energy is negligible.

Much more important is the inclusion of additional bath sites. While the reference system fig.~\ref{fig:fig2} (A) with a single bath site (per correlated site) at the Fermi edge is expected to favor the metallic state, a slight bias towards the insulator is given with reference system (B) where two bath sites are taken into account at energies shifted by $\pm \Delta \varepsilon$ away from the chemical potential $\mu = U/2$. In case of (B), $\Delta \varepsilon$ and $V$ are considered as independent variational parameters.

\section{Results and discussion}

\begin{figure}[t]
\centering \includegraphics[width=0.99\columnwidth]{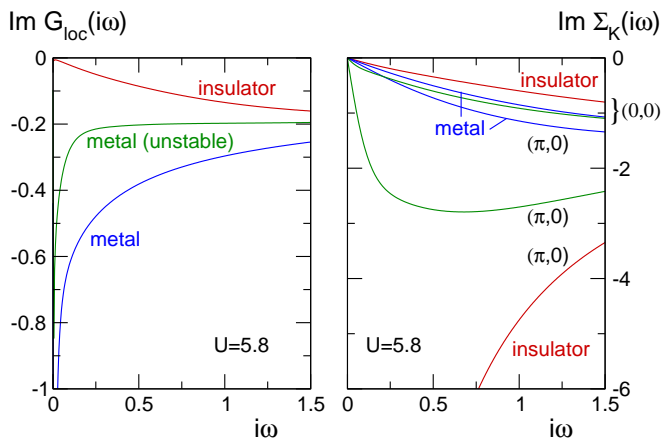}
\caption{ 
Imaginary part of the local Green's function $G_{ii}$ (left) and
of the self-energy $\Sigma_K$ (right) at the cluster momenta $K=(0,0)$ and 
$K=(\protect\pi,0)$ as functions of imaginary frequencies $i\protect\omega$
at $U=5.8$, i.e.\ in the coexistence regime, for the metallic (blue), the
insulating (red) and the third phase (green). Results are obtained
using reference system (A). Note that $\mbox{Im}\Sigma_{(\protect\pi,0)} =
\mbox{Im}\Sigma_{(0,\protect\pi)}$ and $\mbox{Im}
\Sigma_{(\protect\pi,\protect\pi)} = \mbox{Im}\Sigma_{(0,0)}$ due to particle-hole symmetry.
}
\label{fig:fig4}
\end{figure}

We first concentrate on reference system (A).
Fig.~\ref{fig:fig3} shows the dependence of the SFT functional on $V$. 
For $U < U_{c2} \approx 6.35$ (in units of $t\equiv 1$) we find a metallic phase with a
comparatively large optimal hybridization $V_{\mathrm{opt}}$ which decreases
with increasing $U$. For $U > U_{c1} \approx 4.6$ there is a stationary
point of the functional with a much lower $V_{\mathrm{opt}}$ which is less 
$U$ dependent. This corresponds to the Mott insulating phase as is obvious
from the local Green's function and the self-energy displayed in 
fig.~\ref{fig:fig4} for an interaction $U=5.8$ in the coexistence region 
$U_{c1} < U < U_{c2}$: After Fourier transformation of the self-energy on the cluster,
we find the self-energy for $K=(0,0)$ to be regular while for the cluster
momentum $K=(\pi,0)$ it develops a pole at $\omega=0$ (fig.~\ref{fig:fig4},
right). This leads to a vanishing local Green's function $G_{ii}(i\omega)$
for $\omega \to 0$ in the insulating phase (fig.~\ref{fig:fig4}, left). The
same qualitative behavior has been seen at finite $T$ in a recent C-DMFT
study \cite{PHK08}. The metallic solution at $U=5.8$ is characterized by a
finite $G_{ii}(i\omega)$ and a regular $\Sigma_K(i\omega)$ for $\omega \to 0$
(fig.~\ref{fig:fig4}). Note that the $K$ dependence is much weaker in the
metallic solution.

Comparing the ground-state energies at the respective stationary points
(fig.~\ref{fig:fig3}), we find a \emph{discontinuous} metal-insulator
transition at a critical value $U_c=5.79$. The same picture and almost the
same value for $U_c$ (within less than 0.1\%) is found for the simultaneous
and independent three-parameter ($V,t^{\prime},t^{\prime\prime}$)
optimization. Also a different tiling of the square lattice (still using
four-site plaquettes) or using a cluster with two correlated and two bath
sites only (thereby breaking rotational symmetry) does not yield a
qualitatively different picture.

The critical $U_c$ from our plaquette VCA is substantially
smaller than from single-site DMFT. The values $U_c^{\mathrm{(DMFT)}}
\approx 11$ (ref.\ \cite{ZI07}) and $U_c^{\mathrm{(DMFT)}} \approx 12$
(ref.\ \cite{GWTM08}) compare well with our mean-field result 
$U_c^{\mathrm{(DIA)}} \approx 11.3$ which is obtained within the SFT by embedding a single
correlated site coupled to a single bath site into the square lattice. Note
that this two-site dynamical impurity approximation (DIA) is known 
\cite{Pot03b} to reproduce the DMFT phase diagram topology (fig.~\ref{fig:fig1},
top left). We have also verified numerically that it exhibits the
entropy problem. Contrary, the plaquette VCA yields a vanishing ground-state
entropy for (the metal and for) the Mott insulator and consequently supports
quite a different picture, namely (fig.~\ref{fig:fig1}, bottom),
a first-order transition at $T=0$.

Fig.~\ref{fig:fig3} also demonstrates the presence of a third stationary
point (green) in the coexistence region which smoothly links the metal to
the insulator but represents a metastable phase because of its higher
ground-state energy. Note that in VCA, all stationary points are acceptable
solutions. As can be verified from the finite quasiparticle 
weight and the finite density of states at the Fermi edge,
this third phase is metallic in the \emph{entire} coexistence region. This
implies that for four bath sites, there is a quantum
critical point at $U_{c1}$ which marks a \emph{continuous} metal-insulator
transition.
The additional stationary point is reminiscent of the third
solution which is found \cite{TSP01} in single-site DMFT \emph{at finite
temperatures} and which also interpolates between the two main phases.
At finite $T$, however, this is more like a gradual crossover opposed to the 
(hidden) quantum critical point obtained here.

Formally, the plaquette VCA for a single-band model becomes equivalent with
the \emph{single-site} DIA applied to a model with \emph{four orbitals} per
site, namely if the four cluster momenta are identified with the four
orbitals. Contrary to the doped system \cite{FCD+08}, the continuous
transition at $U_{c1}$ is not orbital selective (in this interpretation),
i.e.\ with decreasing $U$ all four $K$-points \emph{simultaneously} undergo
the transition to the insulator (develop a gap in the $K$-dependent spectral
function). The complicated
structure of the Coulomb interaction in $K$ space, and the presence of
correlated-hopping (inter-orbital) terms in particular, makes orbital
selectivity implausible. For $K=(\pi,0)$ and $K=(0,\pi)$ the transition is
of the Mott-Hubbard type. This is consistent with the fact that 
for the particle-hole symmetric case and in the Mott-insulating state, 
the self-energy at $\omega=0$ diverges on the {\em non-interacting} Fermi surface \cite{SPC07}.
For $K=(0,0)$ and $K=(\pi,\pi)$, the insulating spectral function is
neither Mott-Hubbard-like (as it is not particle-hole symmetric) nor
band-insulator-like (as the ``orbital'' $K=(0,0)$ is not fully occupied and
the ``orbital'' $K=(\pi,\pi)$ not completely empty). In the context of
multi-orbital DMFT, a transition to an insulating state with \emph{almost}
complete orbital polarization has been discussed for the titanates \cite{PBP+04}.

\begin{figure}[t]
\centering \includegraphics[width=0.9\columnwidth]{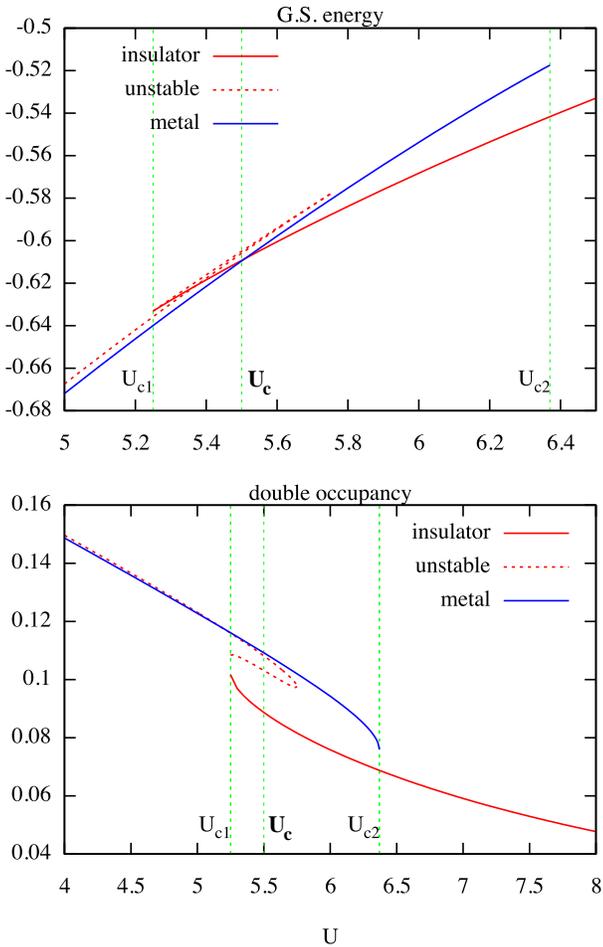}
\caption{
$U$ dependence of the ground-state
energy $E_0$ (top panel) and the
double occupancy $\langle n_{i\uparrow} n_{i\downarrow} \rangle$
(bottom panel) in the metallic (red) and in the insulating (blue)
phase with reference system (B).
The two solutions are not connected; however, the insulating solution
is smoothly connected to an unstable, and then metallic solution
(dashed curves) that together form a pattern similar to the one
observed with reference system (A).
That metallic solution has however a higher energy than the
disconnected metal solution (blue).
The latter is also connected to an unstable solution (not shown) very
close to it. The coexistence range, $U_{c1}$ and $U_{c2}$, is indicated.
The actual transition takes places at $U_c$ where the energies of the two solutions cross.
}
\label{fig:fig6}
\end{figure}

We now move to the reference system fig.~\ref{fig:fig2} (B) with two bath sites per correlated site. Again one finds a first order transition, but with different critical values for the interaction strengths.
To be more specific, fig.\ \ref{fig:fig6} shows the ground-state energy and the double occupancy. There are two sets of disconnected solutions.
Since the metallic and insulating solutions with respective lowest energy belong to families that are not directly connected, it becomes meaningless to try to find the analog of the hidden critical point obtained for reference system (A). 
 
As compared to (A), the coexistence range shrinks. We define the coexistence range from hysteresis. In other words, decreasing $U$ starting from large $U$, we follow the insulating solution (red line) until it disappears. This defines $U_{c1} = 5.25$. We then follow the metallic solution (blue line), increasing $U$ from small $U$, until it disappears at $U_{c2} = 6.37$. These two numbers are more clearly seen on the double-occupancy plot on the bottom panel of fig.\ \ref{fig:fig6}.
The difference $\Delta U_{\rm coex.} = U_{c2}-U_{c1} = 1.12$ is not very different from the $T=0$ extrapolation of the finite-$T$ C-DMFT results of ref.\ \cite{PHK08} ($\Delta U_{\rm coex.} \approx 0.73$) and clearly improves the result $\Delta U_{\rm coex.} = 1.75$ obtained with (A).
This suggests convergence with respect to the number of bath sites similar to the single-site DIA \cite{Poz04}. Note, however, that our results differ from those of C-DMFT solved by exact diagonalization at $T=0$ \cite{ZI07,KT08}.

The critical $U_c$ for the actual metal-insulator transition can be read off from fig.\ \ref{fig:fig6} as $U_c = 5.5$ to be compared with $U_c = 5.8$ obtained with (A).
As could be expected from previous cluster studies \cite{BHP08}, the reference system (B) favors the insulating phase and brings $U_c$ closer to $U_{c1}$ as compared to (A). 
A reference system with three bath sites would favor the metallic phase again.
We therefore believe that the transition would remain first order even if more bath sites were added.

Within DMFT there is an insulating solution above {\em and} below $U_{c2}$ from which at $U_{c2}$ a metallic solution splits off when decreasing $U$.
This is a bifurcation of the insulating solution of the non-linear DMFT equation.
By contrast, within the plaquette approach such a bifurcation mechanism is not necessary since the metal and the insulator have the same (vanishing) ground-state entropy.

\section{Discussion and conclusion}

The Mott insulator is best characterized at high temperature by two properties: a) insulating behavior, i.e., a gap or pseudogap between two peaks in the single-particle density of states, b) {\em no} long-range order~\cite{KHDT04}. 
In the Mott insulator at fixed high $T$, if one decreases $U$, one crosses over to a metallic-like state where the density of states has a single peak. 
One expects the crossover to be replaced by a phase transition at low temperature. 
In reality, long-range antiferromagnetic order is present in the ground state of the Hubbard model with nearest-neighbor hopping only. At infinitesimal $U$ it is driven by nesting. In other words, at small $U$ the gap originates from Slater physics where antiferromagnetic correlations increase with $U$ and the gap increases as $\exp{(-2\pi\sqrt{(t/U)})}$.
At very large $U$, one is in the Heisenberg limit where the antiferromagnetic correlations {\em decrease} with $U$ and the gap {\em increases} linearly with $U$, so Mott physics is relevant. And there is no phase transition between these two limiting antiferromagnets. A phase transition occurs at low enough temperature if antiferromagnetic long-range order is prohibited. In single-site DMFT, one cuts off the correlation length to zero and there is a $T=0$ second-order transition at $U$ about 1.5 times the bandwidth. If one lets the antiferromagnetic correlations grow, the longer the range of these correlations, the closer the transition will be to $U=0$. In the case we studied, antiferromagnetic correlations are cutoff beyond second neighbor. We suggest that this case is the one that is most closely connected to the high-temperature crossover that characterizes a Mott insulator. Indeed, the value $U_c = 5.5$ that we find for the first order transition is closest to the one extrapolated from the finite temperature crossover studied in quantum Monte-Carlo calculations on finite lattices with up to $8\times 8$ sites \cite{VW93}.

At finite $T$, first order transitions are best understood as a tradeoff between energy and entropy in different phases. At $T=0$, we should consider the tradoff between potential and kinetic energy in various phases as the mechanism for the transition. An insulating phase, whether it is stable or metastable, always has lower potential energy than a metallic phase. However, between $U_{c1}$ and $U_c$ the lower kinetic energy (less localization) of the metal makes it more stable than the metastable insulator despite the potential energy advantage of the insulator. Between $U_c$ and $U_{c2}$, the kinetic energy of the metastable metal is not small enough compared with that of the insulator to overcome its lower potential energy. Indeed, the kinetic energy of the insulator can at best grow like $-t^2/U$ in the large $U$ limit while the kinetic energy of the metal can grow faster as electrons are scattered further away from the Fermi surface by the interaction. In practice, the first order transition that we found is in the intermediate coupling regime where the $T=0$ insulator is neither clearly in the Heisenberg (Mott) limit nor in the Slater limit.

The central idea of dynamical (cluster) mean-field theory in general is that
the analysis of the quantum critical point in the paramagnetic
state provides the key to an understanding of the entire phase diagram --
even if this point is obscured by long-range magnetic order: Namely, a $T=0$
critical point of a non-magnetic origin will also rule the physics above the
ordering temperature, in the doped system or in the presence of magnetic
frustration. In the past, this concept has been frequently used at the
single-site DMFT level.

Our main result, however, is that such a quantum critical point is absent. 
The $T=0$ transition is \emph{first order}.
The absence of a continuous non-magnetic $T=0$ transition is contrary to the
widespread DMFT result and therefore expected to have profound
consequences for our understanding of (doped) Mott insulators in two
dimensions. It has been argued based on Hartree-Fock calculations
that explicitly introducing frustration through longer-range hopping should
restore quantum critical behavior at finite frustration \cite{MYI06}, but
that remains to be verified.

\acknowledgments We would like to thank A. Georges, W. Hanke, G. Kotliar, A.
Lichtenstein and A. Liebsch for instructive discussions. Some
numerical calculations were performed on the Sherbrooke Elix4 cluster. The
work is supported by the DFG within the Sonderforschungsbereich SFB~668
(A14) and the Forschergruppe FOR~538 (P1) and by the Natural
Sciences and Engineering Research Council, (Canada), The Canadian Foundation
for Innovation, the Tier I Canada Research chair Program (A.-M.S.T.) and the
Canadian Institute for Advanced Research.

\end{document}